# An Efficient I/O Architecture for RAM-based Content-Addressable Memory on FPGA

Xuan-Thuan Nguyen*, Trong-Thuc Hoang†, Hong-Thu Nguyen†, Katsumi Inoue†, and Cong-Kha Pham†

*Abstract*—Despite the impressive search rate of one key per clock cycle, the update stage of a random-access-memory-based content-addressable-memory (RAM-based CAM) always suffers high latency. Two primary causes of such latency include: (1) the compulsory erasing stage along with the writing stage and (2) the major difference in data width between the RAM-based CAM (e.g., 8-bit width) and the modern systems (e.g., 256-bit width). This brief, therefore, proposes an efficient input/output (I/O) architecture of RAM-based binary CAM (RCAM) for low-latency update. To achieve this goal, three techniques, namely centralized erase RAM, bit-sliced, and hierarchical-partitioning, are proposed to eliminate the latency of erasing stage, as well as to allow RCAM to exploit the bandwidth of modern systems effectively. Several RCAMs, whose data width ranges from 8 bits to 64 bits, were integrated into a 256-bit system for the evaluation. The experimental results in an Intel Arria V 5ASTFD5 FPGA prove that at 100 MHz, the proposed designs achieve at least 9.6 times higher I/O efficiency as compared to the traditional RCAM.

## I. INTRODUCTION

CONTENT-addressable memory (CAM) is a particular kind of memory that allows users to search data by the content instead of the address. It is capable of performing a parallel search over all stored words, find out the match positions, or the addresses, within a single clock cycle. A CAM is classified into a binary CAM and a ternary CAM (TCAM) according to its support bits, i.e., zero, one, and don't care. Due to this parallel-search capability, CAM is applied in a wide variety of applications, such as digital signal processing, data analytics, and pattern recognition [1].

With the significant advances in silicon technologies, modern field-programmable gate arrays (FPGAs) get increasingly favorable because they can produce not only the dedicated-hardware-like performance but also software-like reconfigurability. For example, the 14-nm Intel Stratix 10 FPGA and the 16-nm Xilinx Virtex UltraSCALE FPGA now provide tens of millions of configurable logic and embedded memory blocks that all operate at high clock rate. More significantly, both FPGAs allow users to either partially or entirely reconfigure on-the-fly depending on the critical demands for different tasks. For those reasons, FPGAs have become increasingly important in many emerging systems.

The implementation of CAM in FPGAs have attracted considerable research recently. Unlike the dedicated-hardware CAM made by the custom-designed cells, CAM in modern FPGAs is constructed from the embedded random-access memory (RAM) blocks with a special mapping technique applied in the input data and address [2]-[3]. This specific RAM-based CAM offers the search rate of one key per clock cycle, or O(*1*) processing time.

Recent studies on RAM-based CAM can be categorized into three primary groups including: (1) take advantage of this CAM for parallel search applications [4]-[5], (2) enhance the mapping technique for low-resource utilization [6]-[10], and (3) reduce the update latency for high-input/output (I/O) efficiency [5], [11]. The update stage of RAM-based CAM is composed of two sub-processes, i.e., erase the old contents and write the new ones. It is a crucial factor for high-performance applications because in the traditional RAM-based CAM, the update latency costs O(*2N*) (where $N$ is the number of CAM words), in contrast to the lookup latency of O(*1*). As $N$ increases, the whole performance of RAM-based CAM severely goes down due to the time-consuming update stage. This brief especially focuses on the RAM-based binary CAM (RCAM) and proposes three techniques to minimize its update latency:

- A centralized erase RAM technique that separates the data flows to and from the erase RAM and RCAM, thereby allowing them to operate in parallel.
- A bit-sliced technique that allows RCAM to receive multiple words per clock cycles, thereby fully exploiting the external memory bandwidth.
- A horizontal-partitioning technique that eliminates the latency of both erasing and writing stages, thereby halving the total update latency.

Although the bit-sliced technique was originally introduced in our previous work [5], its full description was not provided yet. One of this work's contributions, therefore, is to expand further details of this useful technique.

The proposed RCAMs at different word sizes, from 8-bit to 64-bit width, are implemented in an Intel Arria V 5ASTFD5K3F40I3 FPGA. The RCAMs access the external DDR3 memory using a 256-bit-bus-width direct memory access controller operating at 100 MHz. The theoretical bandwidth, therefore, reaches as high as 25.6 Gbps. The I/O

Manuscript received ******** ** ****; revised ******** ** ****; accepted ******** ** ****. Date of publication ******** ** ****; date of current version ******** ** ****.
*The author is with the Department of Electrical and Computer Engineering, University of Toronto, Toronto ON M5S 3H7, Canada (e-mail: xuanthuan.nguyen@utoronto.ca).
†The authors are with the Department of Network Engineering and Informatics, the University of Electro-Communications, Tokyo 182-8585, Japan (e-mail: phamck@uec.ac.jp)
Color versions of one or more of the figures in this brief are available online at http://ieeexplorer.ieee.org.
Digital Object Identifier

efficiency, which is defined as a percentage of the update throughput to the DDR3 bandwidth, is deployed to draw a comparison across the other RCAMs. The experiments proved that by using all three techniques above, our RCAMs achieve the update throughput of 24.8 Gbps, or 96.8% of I/O efficiency, regardless of the RCAM width. This result is 9.6 times as high as that of the traditional 64-bit RCAM.

The remainder of this paper is organized as follows. Section II briefly summarizes related works of RCAM. Section III describes the architecture of both traditional RCAM and advanced RCAM using three proposed techniques. Section IV shows the memory utilization and I/O efficiency of our designs in comparison with the traditional one. Finally, Section V presents our conclusion.

## II. RELATED WORKS

Z. Ullah *et al.* [6] presented a hybrid partitioned RAM-based TCAM (HP-TCAM) that first dissected the conventional TCAM table into "m×n" number of TCAM sub-tables. All of the sub-tables were then processed to be stored in their corresponding RAM units. The authors also conducted two alternative architectures, namely E-TCAM [7] and Z-TCAM [8], to bring down the lookup latency as well as the resource utilization.

A. Ahmed *et al.* [9] proposed another memory architecture called resource-efficient RAM-based TCAM (REST), which was partially based on the principle of HP-TCAM. This design could emulate TCAM functionality using more optimal resources, i.e., 3.5% and 25.3% of memory resources compared with HP-TCAM and Z-TCAM, respectively.

W. Jiang [10] also introduced an improved partition technique that could divide a large-sized TCAM into a grid of TCAM units and then assigned each of units to RAM. The number of rows and columns of TCAM grid and the number of words and stages in each TCAM unit could be adjusted to achieve the highest throughput and lowest resource.

F. Syed *et al.* [11] described a fast content-update-module in an HP-TCAM that consumed least possible clock cycles to update a TCAM word. The update latency depends on the number of ternary bits per the selected sub-word. However, the update stage did not take into account the erasing stage. Additionally, the worst latency of writing stage remained O($N$), where $N$ is the number of TCAM words.

D. H. Le *et al.* [4] presented a fast RCAM-based information detection system that could detect each 32×32-pixel template in a 128×128-pixel grayscale image within 61 $\mu$s at a 50-MHz operating frequency. However, such performance was only calculated based on the lookup latency.

Our previous work [5] presented a 10.2-Gbps bitmap-index-based regular expression matching based on RCAM architecture. The bit-sliced technique was employed to dramatically reduce the update latency. However, its detailed description was not provided. Furthermore, the problem of erasing stage was still not resolved.

## III. IMPLEMENTATION

### A. Traditional Design

A 32×8-bit RCAM, or RCAM unit (RCU), which is made by a dual-port memory, is illustrated in Fig. 1 (right-side module). Port A is seen as an 8,192×1-bit memory (13-bit address and 1-bit data), while port B is seen as a 256×32-bit memory (8-bit address and 32-bit data). With this setting, each RCU is able to fit into a basic embedded memory unit called M10K block of an Intel Arria V FPGA. The input data ($cdata$) and address ($caddr$) enter port A, while the search key ($ckey$) and matching address ($cmatch$) come out from port B. Before receiving new data, RCAM has to clear the stored values by setting the write enable ($cwe$) to one and set/clear ($csc$) to zero. New input data are then written by asserting $csc$ in the following cycle. In other words, the erasing and writing stages are compulsory to updating RCAM. Full details of RCAM operation can be found in [2]-[3].

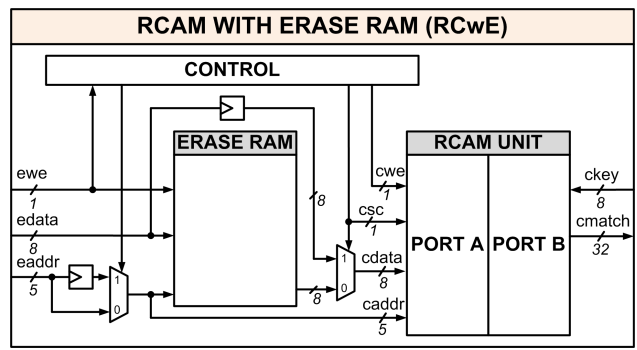

Fig. 1: The block diagram of a traditional 32×8-bit RCwE.

As a result of the two-stage update process, an erase RAM is added ahead of RCU to form a so-called RCwE core [3], as depicted in Fig. 1. All input signals including $ewe$, $edata$, and $eaddr$ pass through the erase RAM. A control module, together with two multiplexers and two buffer registers, are deployed to manage both erasing and writing stages. At the first clock cycle (i.e., erasing stage), the control module assigns $cwe = 1$ and $csc = 0$ so that a RCAM cell is erased. It is noted that the data at $eaddr$ is the last data stored in the RCU. At the second clock cycle (i.e., writing stage), the data at $eaddr$ is replaced by the new $edata$. Simultaneously, the control module asserts $csc$ so that new $edata$ is put into RCU. In short, each update stage consists of two interleaved erasing and writing stages, which costs two clock cycles in total.

Both CAM depth and width can be simply expanded by connecting a set of RCwEs according to the specific models. An example of depth expansion, where the number of RCAM cells increases from 32 to 64, is shown in Fig. 2(a). The input data are put into each sub-RCwE in turn, from RCwE$_0$ to RCwE$_1$. As soon as RCwE$_0$ is full, RCwE$_1$ starts receiving data. The process repeats until RCwE$_1$ is completely filled. An example of width expansion, where the RCAM width increases from 8 to 16 bits, is depicted in Fig. 2(b). Unlike the previous architecture, the input data are divided into two segments, and each of which enters each correspondent sub-RCwE simultaneously. The output can be seen as 32 two-bit groups and each of which connects to an AND gate. This is because a match position is defined as all group bits matching the read input on the same address. Finally, both depth and width expansion is obtained by combining two architectures above, as seen in Fig. 2(c).

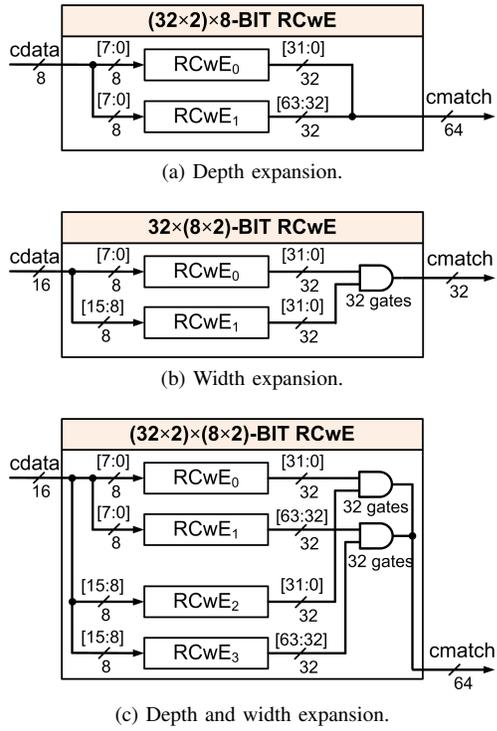

Fig. 2: The cascade architecture of a traditional RCwE.

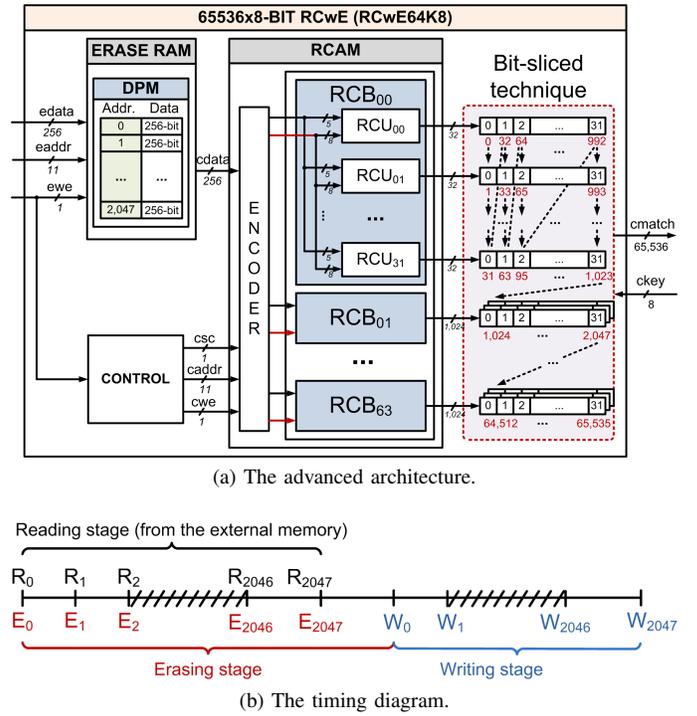

Fig. 3: The advanced architecture and timing diagram of RCwE64K8 using centralized erase RAM and bit-sliced techniques.

The main disadvantage of the traditional RCAM, as well as the previous works [4], [6]-[11], is the update latency. Suppose that $N$ is the number of CAM words, the total update latency becomes O($2N$) because of the erasing and writing stages. In comparison with the search time O($1$), the update time slows down the whole design. Another disadvantage of the traditional RCAM is the huge number of memory usage. Because each RCwE consists of one erase RAM and one RCU, and each of which costs an M10K block, the total used M10K blocks is doubled. Moreover, each M10K block can store up to eight Kbits, whereas each erase RAM only contains 32 8-bit RCU values, or 256 bits in total. This implies that only 3.2% (256/8,192) of M10K block is used for the erase RAM, or as high as 96.8% of that M10K capacity is wasted.

*B. Advanced Design Using Centralized Erase RAM and Bit-Sliced Techniques*

Fig. 3(a) illustrates the architecture of an advanced 65,536×8-bit RCwE (RCwE64K8) designed for a 256-bit system. To fully exploit the throughput of such data width, we first partition RCAM into 64 sub-RCAM blocks (RCBs), each of which consists of 32 RCUs. With this partition, RCwE64K8 requires 2,048 RCUs. In the update stage, each 256-bit $cdata$ are put into 32 RCUs of RCB$_{00}$ simultaneously. As soon as RCB$_{00}$ is full, $cdata$ is navigated to RCB$_{01}$ owing to the encoder module. This process repeats until the entire RCBs are full.

Subsequently, the bit-sliced technique is applied in RCwE64K8 output to guarantee the correct order. Specifically, the first 32 bits of $cmatch$ are formed by 32 zeroth bits of 32 RCUs inside RCB$_{00}$; the second 32 bits of $cmatch$ are built by 32 first bits of all RCUs inside RCB$_{00}$, and so on. The first 1-Kbit is constructed by sequentially connecting 32 32-bit segments created. The remaining RCBs follow the same procedure. As a result of the bit-sliced technique, RCAM is capable of receiving as high as 32 8-bit words concurrently, whereas the order of its 64-Kbit output is kept unchanged. Accordingly, the write latency of RCwE64K8 reduces up to 32 times as compared to the traditional design.

Furthermore, all of the individuals erase RAMs are centralized into a single 2,048×256-bit dual-port memory (DPM) to resolve the memory utilization. This DPM is only built by 64 M10K blocks, instead of 2,048 M10K blocks as of the traditional RCwE. Fig. 3(b) shows the schedule of erasing and writing stages. In the beginning, if $ewe$ is asserted, new $edata$ is copied into erase RAM at $eaddr$. Simultaneously, the control module asserts $cwe$ together with $csc$ to clear the CAM using $cdata$ and $caddr$. Due to the centralized erase RAM, both erasing and writing stages can run separately. After fully erasing RCAM, the control module deasserts $csc$ so that RCAM receives new data. Finally, $cwe$ is deasserted to end the updating process. The total latency, hence, drops from O($2N$) to O($\frac{2N}{32}$). Overall, if the number of RCUs per RCBs is defined as $k$ (i.e., $k = 32$ in this scenario), then the latency can be expressed as O($\frac{2N}{k}$).

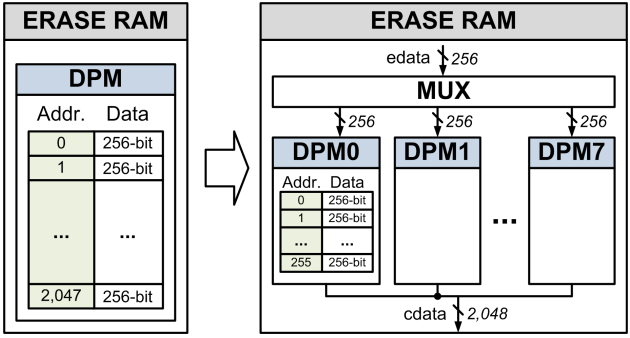

Fig. 4: The illustration of horizontal-partitioning technique.

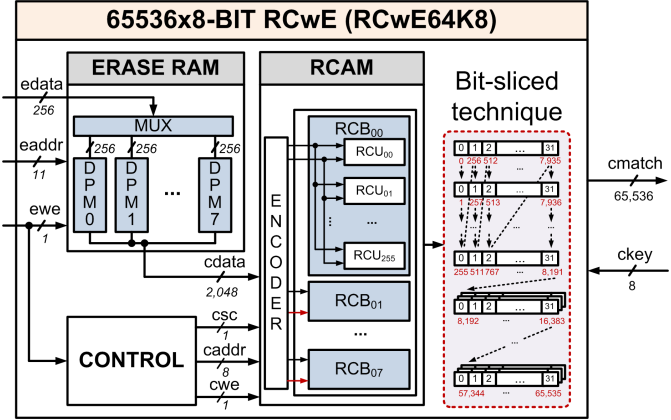

(a) The advanced architecture.

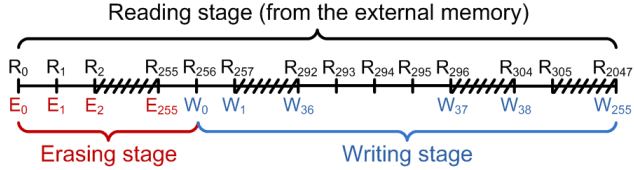

(b) The timing diagram.

Fig. 5: The advanced architecture and timing diagram of RCwE64K8 using all of three techniques.

*C. Advanced Design Using Centralized Erase RAM, Bit-Sliced, and Horizontal-Partitioning Techniques*

As seen in Fig. 3(b), the read latency from the external memory is limited by the 256-bit system width. However, it is possible to expand the data width of $cadata$ in order to dramatically shrink the latency of both erasing and writing stages. Based on this concept, we introduce a so-called horizontal-partitioning technique, which is demonstrated in Fig. 4. Instead of constructing a large-sized erase RAM, we divide it into eight sub-modules arranged horizontally. Each sub-module is made from a 256×256-bit DPM. A multiplexer is then used to select a certain sub-module to receive $edata$. Specifically, eight consecutive $edata$ values are put into DPM0[0] to DPM7[0] during the first eight clock cycles. The rest of $edata$ are continued loading similarly until all of the sub-modules are full. By contrast with receiving 256-bit data, as high as 2,048-bit $cdata$ is get out in every clock cycle.

Fig. 5(a) gives the advanced architecture of RCwE64K8

| RCwE size (bits) | Adaptive logic modules | Embedded M10K blocks | Frequency (MHz) |
|---|---|---|---|
| 65,536×8 | 109 (0.06%) | 2,112 (74%) | 134.2 |
| 32,768×16 | 16,773 (9.5%) | 2,112 (74%) | 134.9 |
| 16,384×24 | 16,617 (9.4%) | 2,112 (74%) | 135.3 |
| 8,192×32 | 19,017 (10.8%) | 2,112 (74%) | 135.2 |

TABLE I: The place-and-route results.

employing centralized erase RAM, bit-sliced, and horizontal-partitioning techniques. Since the current $cdata$ signal contains 2 Kbits, the number of RCBs reduces to eight modules, whereas $k$—the number of RCUs per RCB—increases to 256 modules.

Fig. 5(b) depicts the new timing diagram of update stage. Suppose that there is no latency in the external memory access, 2,048 clock cycles are spent to copy all 64-KB data to the erase RAM through the 256-bit bus width. However, only 256 clock cycles are required to entirely erase the RCAM due to the eight times larger in the bus width. As a result, the writing stage can be started upon the $255^{th}$ clock cycle. After the next 36 clock cycles, RCAM receives all new data available in erase RAM. From this point, as soon as all of eight DPMs' rows are properly filled by the 256-bit $edata$, a new 2-Kbit $cdata$ is dispatched to RCAM. By using this mechanism, the writing stage is finished at the same time with the read stage. In other words, the compulsory erasing stage no longer affects the whole latency of update stage. The total updating latency, therefore, goes down from $O(\frac{2N}{k})$ to $O(\frac{N}{k})$.

IV. PERFORMANCE ANALYSIS

This section proves the advantages of the propose RCwE against the traditional one regarding the resource utilization and I/O efficiency.

Table I states the resource utilization of a RCwE64K8, a 32,768×16-bit RCwE, a 16,384×32-bit RCwE, and a 8,192×64-bit RCwE that were all implemented in an Intel Arria V 5ASTFD5K3F40I3 FPGA. The design software is Quartus Prime 16.0, and the compiler setting is aggressive performance mode. The results of adaptive logic modules, embedded M10K blocks, and operating frequency are used to evaluate those designs. Each logic module consists of one 8-input combinational look-up table with four dedicated registers and is seen as a fundamental building logic block of an Arria V FPGA. As shown in Table I, RCwE64K8 consumed almost none of the logic modules, whereas the others cost approximately 10% of total logic modules because of a large number of AND gates discussed in Fig. 2(b). The number of M10Ks, however, remain unchanged due to the same number of used RCUs. The maximum operating frequencies are reported by the place-and-route timing analysis.

Fig. 6 depicts the number of M10Ks used by the traditional RCwE in Section III-A (RCwE_S1 [2]-[3]) and two alternative advanced RCwEs in Section III-B (RCwE_S2) and Section III-C (RCwE_S3). In RCwE_S1 design, RCAM and erase RAM share 50% of the total 4,096 M10K blocks. However, both RCwE_S2 and RCwE_S3 only require 2,112 M10K blocks, i.e., 2,048 blocks for RCAM and 64 blocks for erase

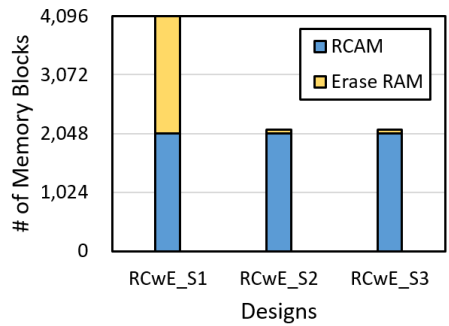

Fig. 6: The memory utilization.

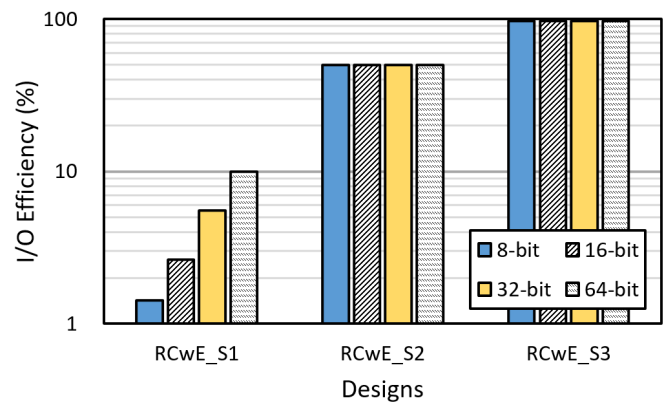

Fig. 7: The I/O efficiency.

RAM. In other words, both advanced RCwEs saved 48.4% of M10K blocks compared to the traditional one.

In order to measure the I/O efficiency, a test system containing a direct memory access controller and an advanced RCwE was designed. This controller allowed RCwE to access data directly in a 1-GB onboard DDR3 memory. Because this external memory provided a theoretical bandwidth of 25.6 Gbps, the controller's bus width and operating frequency were configured as 256 bits and 100 MHz, respectively. Fig. 7 gave the I/O efficiency of RCwE_S1, RCwE_S2, and RCwE_S3 when the RCAM width ranged from 8 bits to 64 bits.

• In RCwE_S1, a shift register was inserted to convert a 256-bit input data into the corresponding CAM widths. For example, in case of 8-bit input data, RCwE_S1 needed 32 clock cycles to fully receive a certain input. The I/O efficiency of RCwE_S1, therefore, tended to increase corresponding to the RCAM width. In case of maximum 64-bit bus width, the throughput reached 2.6 Gbps, or approximately 10.1% of the theoretical bandwidth. All of the previous works [4], [6]-[11] also produced similar I/O efficiency due to the lack of optimization in the update stage.

• In RCwE_S2, by using the centralized erase RAM and bit-sliced techniques, RCwE requested data continuously, thereby benefited greatly from the DDR3 memory access, such as bank interleaving. Moreover, the throughput became stably disregarding the RCAM width. It reached around 12.7 Gbps, or 49.8% of theoretical bandwidth, due to the requirement of the erasing and writing stages.

• In RCwE_S3, by further applying the hierarchical-partitioning technique, both erasing and writing stages could operate in parallel with the read DDR3 stage. Hence, the throughput was boosted to 24.8 Gbps, or around 96.8% of the theoretical bandwidth. This proposed RCwE achieved 9.6 times (96.8/10.1) as high as I/O efficiency and 48.4% as low as memory utilization, as compared to the traditional 64-bit RCwE_S1.

As mentioned earlier, the update latencies of RCwE_S1 and RCwE_S3 are O($2N$) and O($\frac{N}{k}$), respectively. As a result, in case of 64-bit width, the update throughput of RCwE_S3 should become eight times higher than that of RCwE_S1. However, since RCwE_S1 accessed DDR3 memory discretely, it suffered higher latency caused by, for instance, DDR3 precharge and row activation. Therefore, the throughput difference between RCwE_S1 and RCwE_S3 inflated to 9.6 times.

V. CONCLUSION

This brief has presented an I/O efficient architecture of a RAM-based CAM on FPGA. By employing the centralized erase RAM, bit-sliced, and hierarchical-partitioning techniques, we can eliminate the effect of erasing stage on the updating stage and fully exploit the bandwidth of the external memory. The design can be adaptive effectively according to the bus width of external memory and RAM-based CAM core. As compared to the traditional RAM-based CAM, our design achieves at least 9.6 times higher I/O efficiency and 48.3% lower memory utilization.